# Anomalous spin–lattice coupling in a 2D antiferromagnetic semiconductor revealed by surface acoustic Rayleigh waves

Zahra Ebrahim Nataj,[1,2] Md Sabbir Hossen Bijoy[3], Vladislav Korostelev[3], Dylan Wright[1,2], Mohammad Zeinolabedini[3], Konstantin Klyukin[3], Fariborz Kargar [3,*], Alexander A. Balandin[1,2,*]

[1]Department of Materials Science and Engineering, University of California, Los Angeles, California 90095 USA

[2]California NanoSystems Institute, University of California, Los Angeles, California 90095 USA

[3]Materials Research and Education Center, Department of Mechanical Engineering, Auburn University, Auburn, Alabama 36849 USA

* Corresponding author: fkargar@auburn.edu (F.K.); balandin@seas.ucla.edu (A.A.B.)

# Abstract

Magnetic order in van der Waals magnets can strongly influence their lattice dynamics, yet how this interaction manifests across different phonon length scales remains unclear. Optical phonons probe bond-scale exchange modulation and short-range spin correlations, whereas long-wavelength acoustic modes couple to uniform strain fields and are sensitive to the renormalization of the macroscopic elastic tensor associated with long-range magnetic order. Experimentally accessing these low-energy acoustic excitations in low-dimensional crystals is challenging due to their low energies and the small lateral dimensions of exfoliated samples. Here, we employ angle-resolved Brillouin–Mandelstam scattering spectroscopy to investigate the surface acoustic phonon spectrum of exfoliated $NiPS_3$ thin films across their antiferromagnetic transition temperature. Our results show a single Rayleigh surface mode whose phase velocity exhibits a pronounced ~5.5% softening upon cooling through the Néel temperature. This anomaly reflects a giant magnetoelastic renormalization of the long-wavelength elastic constants triggered by the onset of zigzag antiferromagnetic order. First-principles calculations of the full elastic tensor, combined with continuum finite-element modelling of the $NiPS_3/SiO_2/Si$ heterostructure, reproduce both the Rayleigh-wave dispersion and its magnetic-order–induced shift. The obtained results reveal how microscopic exchange interactions shape macroscopic mechanical properties in two-dimensional antiferromagnetic semiconductors, providing a basis for lattice-controlled magnetism and magnetically tunable phononic, magnonic, and strain-mediated spintronic device concepts.

Two-dimensional (2D) van der Waals (vdW) magnets offer a unique platform for probing and controlling the interplay between spin and lattice degrees of freedom at the atomic scale[1–3]. While the Mermin–Wagner theorem prohibits long-range magnetic order in the limit of a single isotropic layer[4], real vdW magnets stabilize finite-temperature order through magnetocrystalline anisotropy arising from spin–orbit coupling, which lifts the degeneracy of spin orientations and opens a gap in the spin-wave spectrum[2,5–8]. The magnitude of this anisotropy is highly sensitive to lattice geometry since the exchange pathways and crystal-field environment depend on interatomic bond lengths and angles[8,9]. Spin–phonon, *i.e.*, magnetoelastic, coupling provides the microscopic link between lattice dynamics and magnetic anisotropy, enabling atomic displacements and strain fields to modulate exchange interactions and crystal-field splitting[8–10]. Lattice distortions – whether static, *e.g.*, strain, pressure, stacking, or dynamic, *i.e.*, phonons – can renormalize the anisotropy and modify the magnetic excitation spectrum[8–12]. The reciprocal effect also holds where the onset of magnetic order reshapes the lattice through magnetostriction[13,14]. In the strict 2D limit, where the magnetic ground state is fragile, even small lattice distortions can strongly influence magnetic stability. Understanding phonon renormalization and elastic softening due to spin–phonon coupling provides insight into how exchange interactions and magnetic anisotropy evolve across the transition and how the lattice participates in stabilizing the ordered phase.

In many ferro- and antiferromagnets, the strength of spin–phonon coupling has traditionally been evaluated by tracking the temperature dependence of Raman-active optical phonons across the magnetic ordering transition[15–19]. In this approach, deviations of the phonon frequency below Néel temperature, $T_N$, from the expected anharmonic temperature trend are attributed to the influence of evolving spin correlations on local bonding environments[19]. Note that optical modes involve out-of-phase atomic motions within the unit cell and are therefore sensitive to short-range exchange pathways, making Raman spectroscopy an effective probe of *local* spin correlations[20]. Microscopically, this arises because the phonon frequency shift depends on the spin–spin correlation function $\langle S_i \cdot S_j \rangle$, which reflects nearest-neighbor exchange interactions and persists even above $T_N$ where only short-range magnetic correlations remain[20]. Optical phonons, therefore, primarily reflect bond-scale exchange modulation and do not directly capture the macroscopic elastic response of the lattice associated with *long-range* magnetic ordering. Acoustic phonons, in contrast, correspond to long-wavelength, in-phase lattice displacements governed by the macroscopic bulk and shear moduli. Their softening or hardening across the magnetic transition reflects the magnetoelastic renormalization of the elastic constants, providing a direct measure of coupling at collective, continuum length scales[21–24]. Despite their distinct importance at larger length scales, experimental studies have predominantly focused on optical phonons, and the behavior of acoustic phonons in vdW magnets remains largely unexplored due to the challenges of probing them in small-size exfoliated samples with conventional neutron or ultrasound methods.

In this work, we investigate how antiferromagnetic (AFM) ordering in $NiPS_3$ influences its long-wavelength elastic response by directly probing surface acoustic phonons in exfoliated thin films on $SiO_2$/Si substrates. Although the influence of magnetic ordering on local bond-scale dynamics in $NiPS_3$ is well established, it remains unclear whether and how this ordering modifies its long-wavelength elastic response. $NiPS_3$ is a quasi-two-dimensional antiferromagnet with XY-type spin anisotropy and dominant intralayer exchange interactions[9,25–27]. The $Ni^{2+}$ spins order into a zigzag pattern below its Néel temperature, $T_N \sim 155\ K$, producing two low-energy magnon gaps and a spin–orbit–entangled excitonic excitation observed in neutron and Raman experiments[28,29]. Prior Raman studies have revealed mode-selective spin–phonon coupling in specific optical modes of $NiPS_3$, including pronounced Fano-like line-shape modifications and mode splitting below its Néel temperature[16,30]. However, the response of low-energy acoustic phonons to AFM ordering in $NiPS_3$ remains unknown. We employ angle-resolved Brillouin–Mandelstam light scattering (BMS) spectroscopy, which allows access to Rayleigh-type surface acoustic modes confined in micron-scale supported films. Tracking these modes across the Néel temperature enables a direct measurement of how the elastic constants evolve with spin ordering. By combining the BMS measurements with first-principles elastic tensor calculations and continuum finite-element modeling, we establish a quantitative link between magnetic order and lattice elasticity in $NiPS_3$. This approach provides a pathway for assessing magnetoelastic coupling in vdW AFM semiconductors and informs the design of phononic and spintronic devices that leverage spin–lattice interactions.

## Results and Discussions

In this study, we used high-quality single crystals of NiPS$_3$ (2D Semiconductors, USA). Thin films were prepared by mechanical exfoliation onto SiO$_2$/Si substrates with a 300 nm oxide layer. Figure 1 (a) shows the bulk crystal structure of NiPS$_3$. The compound crystallizes in the monoclinic space group C2/m. Within each layer, Ni atoms form a honeycomb structure coordinated by P$_2$S$_6$ molecular units[31]. Figure 1(b) illustrates the magnetic ground state. Below the Néel temperature, NiPS$_3$ adopts a zigzag AFM order in which ferromagnetic chains run along the *a*-axis and are antiferromagnetically coupled along the *b*-axis with the magnetic moments lying predominantly within the *ab* plane and having only a small out-of-plane component. Adjacent layers are additionally coupled antiferromagnetically along the *c*-axis. This magnetic ground state is characterized by the spin propagation vector **k** = [010][32].

Understanding the crystallographic orientation of exfoliated films is important for surface acoustic phonon studies, as it determines the probed phonon wavevector direction. Similar to FePS$_3$, NiPS$_3$ exhibits well-defined in-plane cleavage edges corresponding to the <100> or <110> crystallographic directions, which form angles of 60° [Refs. [33,34]]. These edges align with the zigzag (*a*-axis) direction, while the armchair direction (*b*-axis) is perpendicular to it. This allows reliable identification of the lattice axes using optical microscopy. The *a* and *b* axes of the exfoliated films were determined directly from the optical microscopy image, as shown in Supplementary Fig. 1. The thickness of the exfoliated NiPS$_3$ films was determined using atomic force microscopy, which showed an average thickness of ~220 nm (Supplementary Fig. 2).

Raman spectroscopy was performed on the exfoliated NiPS$_3$ films using a 532 nm laser excitation at room temperature (RT) to confirm the sample quality. Bulk NiPS$_3$ belongs to C$_{2h}$ point group and its irreducible Γ-point phonon representation is $\Gamma = 8A_g + 7B_g + 6A_u + 9B_u$, where $A_g$ and $B_g$ modes are Raman active and $A_u$ and $B_u$ modes are infrared active[25,34,35]. Figure 1(c) shows the unpolarized Raman spectrum in which all Raman-active modes are observed, confirming the structural quality of the sample. The low-frequency modes arise mostly from the vibration of the Ni atoms, whereas the higher-frequency modes above originate from the intermolecular vibrations of the $(P_2S_6)^{-4}$ units. As seen in Fig. 1(c), several nearly-degenerate pairs of $A_g$ and $B_g$ modes are observed. These modes originate from the splitting of the monolayer $E_g$ phonons due to the monoclinic stacking in bulk NiPS$_3$, which lifts the degeneracy present in the monolayer[25,35]. To distinguish $A_g$ and $B_g$ symmetries, we conducted polarization-resolved Raman spectroscopy. In backscattering geometry, both $A_g$ and $B_g$ modes are visible in the parallel configuration, $z(xx)\bar{z}$, while only $B_g$ modes are present in the cross-polarized configuration, $z(xy)\bar{z}$, as shown in Figs. 1 (d, e). Lorentzian fitting was applied to all Raman peaks, with the fitted components shown as shaded areas in Figs. 1(c–e). The weak feature marked by an asterisk at ∼450 cm$^{-1}$ may originate from higher-order Raman scattering and is therefore not assigned here. The positions and the full-width-at-half-maximums (FWHM) of the observed peaks are listed in Supplementary Table 1. The obtained Raman peaks are consistent with previously published data[25,34,36].

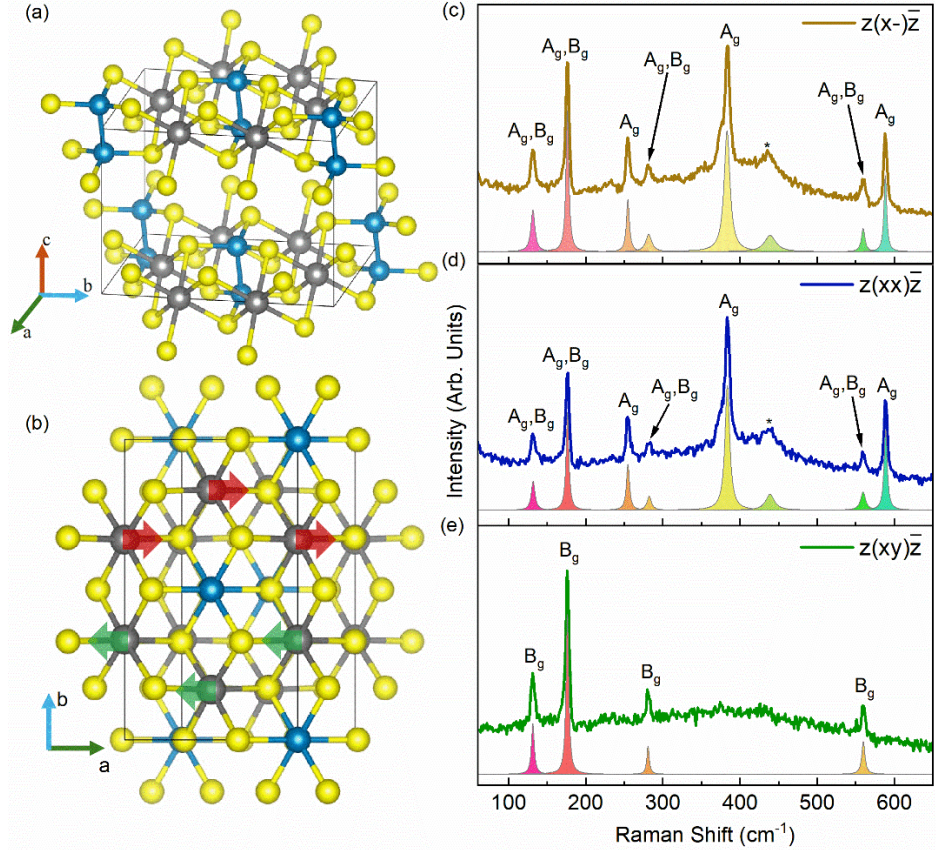

**[Fig. 1: Crystal, magnetic, and vibrational structure of monoclinic NiPS$_3$.** (a) Crystal structure of monoclinic NiPS$_3$. Ni atoms form a honeycomb lattice coordinated by P$_2$S$_6$ units. Ni, P, and S atoms are depicted with black, blue, and yellow spheres, respectively. (b) View of the structure along the *c*-axis. The arrows show the magnetic spin orientation lying predominantly within the *ab* plane. (c-e) Raman spectra collected in unpolarized, $z(x-)\bar{z}$, parallel $z(xx)\bar{z}$, and cross-polarized $z(xy)\bar{z}$ configurations, respectively. The polarization-selective measurements enable the identification of $A_g$ and $B_g$ phonon symmetries. Peaks are fitted using Lorentzian functions shown as shaded areas.]

To assess the conventional spin-phonon coupling, we performed temperature-dependent Raman spectroscopy across $T_N$. Figure 2(a) shows the accumulated Raman spectra in the temperature range of 90 K to 290 K. A slight Fano asymmetry is visible in the ~255 cm$^{-1}$ mode, consistent with earlier reports[16] linking this feature to coupling between phonons and a magnetic excitation continuum[16,30]. Figures 2(b–e) present the temperature dependence of the peak positions and linewidths for two representative modes at 255 cm$^{-1}$ and 588 cm$^{-1}$. As stated previously, the low-frequency modes primarily involve the vibration of Ni atoms, where the magnetic moments are localized[9], whereas the higher frequency modes originate mainly from internal vibrations of P$_2$S$_6$ units. In principle, one should expect the low-frequency modes to be the most sensitive to magnetic ordering. However, the 255 cm$^{-1}$ mode exhibits only a slight and narrow flattening around $T_N$ before resuming its anharmonic behavior (Fig. 2(b)), and its FWHM deviates by only ~0.3 cm$^{-1}$ from the anharmonic trend. The higher-frequency mode at 588 cm$^{-1}$ follows the expected anharmonic temperature evolution throughout the entire examined range with no detectable signature of magnetic ordering. The spectral position and FWHM of other peaks observed in Fig. 2(a) are provided in Supplementary Fig. 3. All observed changes for other peaks remain within fitting and instrumental uncertainty and therefore do not constitute a clear or systematic deviation at the transition. Prior polarization-resolved Raman studies have reported a small splitting of the peak at ~176 cm$^{-1}$ across $T_N$ [25], but our polarization-resolved measurements show no such splitting (Supplementary Fig. 4). These observations suggest that the renormalization of the optical phonon frequencies driven by the onset of long-range magnetic order is small in NiPS$_3$, and that other manifestations of spin-phonon coupling, such as changes in peak line-shape and intensity, are subtle and require higher spectral sensitivity to resolve reliably. Because the optical phonons do not provide a sufficiently sensitive probe of the spin–phonon coupling in NiPS$_3$, we next examine the low-energy surface acoustic phonons using BMS spectroscopy, which directly probes magnetoelastic interactions.

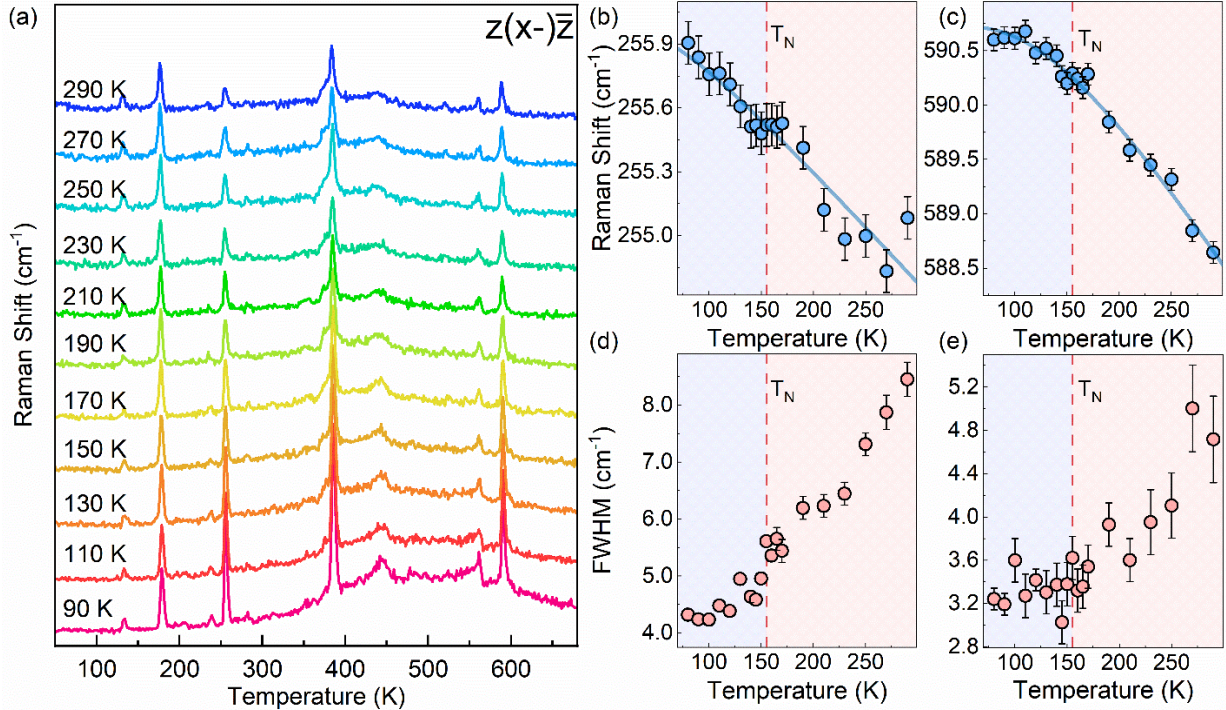

[**Fig. 2: Temperature-dependent evolution of optical phonon modes in NiPS$_3$.** (a) Raman spectra collected in the $z(x-)\bar{z}$ polarization configuration from 90 to 290 K. (b, c) Temperature evolution of the phonon frequencies for the modes centered at ~255 cm$^{-1}$ and ~588 cm$^{-1}$, respectively. The solid curves represent fits to the anharmonic model, and the vertical dashed line marks the Néel temperature, $T_N$. (d, e) Full width at half maximum (FWHM) of the same two modes as a function of temperature. A small deviation from the anharmonic trend is observed near $T_N$ for the 255 cm$^{-1}$ mode, whereas the 588 cm$^{-1}$ mode exhibits no detectable anomaly.]

We performed BMS experiments in the conventional backscattering geometry to probe the low-energy acoustic phonons near the Brillouin-zone center. Similar to Raman spectroscopy, BMS relies on inelastic light scattering, but the energy range accessed is several orders of magnitude lower, corresponding to acoustic phonons with wavevectors close to the Γ-point. Further details of the technique can be found in prior work from our group and others[37–39]. Because the thickness of our exfoliated NiPS$_3$ thin films (~220 nm) is smaller than the BMS laser excitation wavelength, $\lambda = 532$ nm, the BMS spectra are dominated by surface acoustic waves[37,38]. In this thickness regime, the out-of-plane component of the phonon momentum is not uniquely defined, and only the in-plane component of momentum is conserved in the scattering process[38]. The in-plane phonon wavevector is given by $q = (4\pi/\lambda)\sin\theta$, where $\theta$ is the angle of incidence measured relative to the surface normal. To select the crystallographic direction of the probed phonons, the sample was rotated in-plane so that the in-plane phonon wavevector, $q$, was aligned with the sample's *b*-axis, determined independently from optical microscopy (Supplementary Fig. 1). After fixing the in-plane orientation, varying $\theta$ allowed us to tune the magnitude of $q$ and extract the dispersion, *i.e.*, energy *vs.* $q$, of the surface acoustic mode.

The accumulated spectra obtained for $\theta$ ranging from 30° to 50° at RT are shown in Fig. 3 (a). We identified a single well-defined dispersive peak, whose frequency increases with $q$. This peak is attributed to the Rayleigh surface acoustic wave in $NiPS_3$. To verify that the observed mode originates from the $NiPS_3$ layer rather than the underlying $SiO_2$/Si substrate, we also measured the BMS spectra of a bare $SiO_2$/Si substrate under the same experimental geometry (Supplementary Fig. 5). The substrate exhibits multiple features associated with Rayleigh, Sezawa, and Lamb surface acoustic waves, consistent with the literature[40]. These modes exhibit different phase velocities and spectral positions than the single dispersive peak observed in $NiPS_3$, confirming that the latter originates from the Rayleigh wave in the exfoliated $NiPS_3$ layer. The assignment of this mode will be further corroborated through comparison with calculated surface acoustic wave dispersions later.

The spectral position, $f$, of this peak at different $\theta$s was obtained by fitting the experimental data with individual Lorentzian functions (shaded area in Fig. 3(a)). The resulting values are plotted as a function of the corresponding $q$, as shown by blue circles in Fig. 3(b). The phase velocity was then calculated using $v = 2\pi f/q$, and is depicted by red spheres on the same plot. The dispersion of the Rayleigh surface wave in a layered structure depends on whether the film is slower or faster than the substrate with respect to the relevant in–plane acoustic velocities[37,41]. In a slow-on-fast configuration, where the film has a lower in-plane shear acoustic velocity than the substrate, the Rayleigh mode at long wavelengths (small $q$) penetrates into the substrate and therefore exhibits a higher effective phase velocity. As $q$ increases, the acoustic wavelength becomes shorter and the surface wave becomes progressively more confined to the $NiPS_3$ layer, reducing substrate participation. The decreasing phase velocity of the peak with increasing $q$, shown in Fig. 3(b) indicates that, for the phonon wavevector aligned along the *b*-axis, the $NiPS_3$/$SiO_2$/Si heterostructure behaves effectively as a slow-on-fast system for the surface acoustic wave.

To identify the nature of the observed dispersive mode, we calculated the elastic stiffness tensor, $c_{ij}$, of NiPS$_3$ using density-functional theory (DFT) with PBE-D3 correction in both the AFM and non-magnetic (NM) states. Since NiPS$_3$ is paramagnetic (PM) above $T_N$, the NM state is used as an approximation to the high-temperature PM phase. The resulting elastic constants were incorporated into a continuum elastic model of the NiPS$_3$/SiO$_2$/Si heterostructure implemented using finite-element simulations in COMSOL Multiphysics. Additional computational details are provided in the Methods, and validation of the simulation setup is checked with SiO$_2$/Si configuration shown in Supplementary Fig. 6. The eigenmode analysis yields several surface and guided acoustic branches, but only one mode displays surface-localized elliptical displacement predominantly within the NiPS$_3$ layer (Fig. 3(c)), which is the characteristic vibrational signature of a Rayleigh surface acoustic wave. According to Loudon's selection rule for surface Brillouin scattering, only modes with displacement components perpendicular to the phonon propagation direction contribute significantly to the scattered intensity[38,42,43]. Therefore, this Rayleigh mode is expected to dominate the BMS spectrum. The calculated frequency of this mode at $q$=0.0135 nm$^{-1}$ is ~4.6 GHz, which is in close agreement with the experimentally observed value of ~4.2 GHz. We also calculated the dispersion and phase velocity of this surface mode, shown with blue and red circles, respectively, in Fig. 3(d). As seen, the theoretical calculations follow the same decreasing phase-velocity trend observed experimentally in Fig. 3(b). This agreement between experiment and modeling further validates the mode assignment and indicates that the Rayleigh wave becomes increasingly confined to the NiPS$_3$ layer as the in-plane wavevector increases (Supplementary Fig. 7).

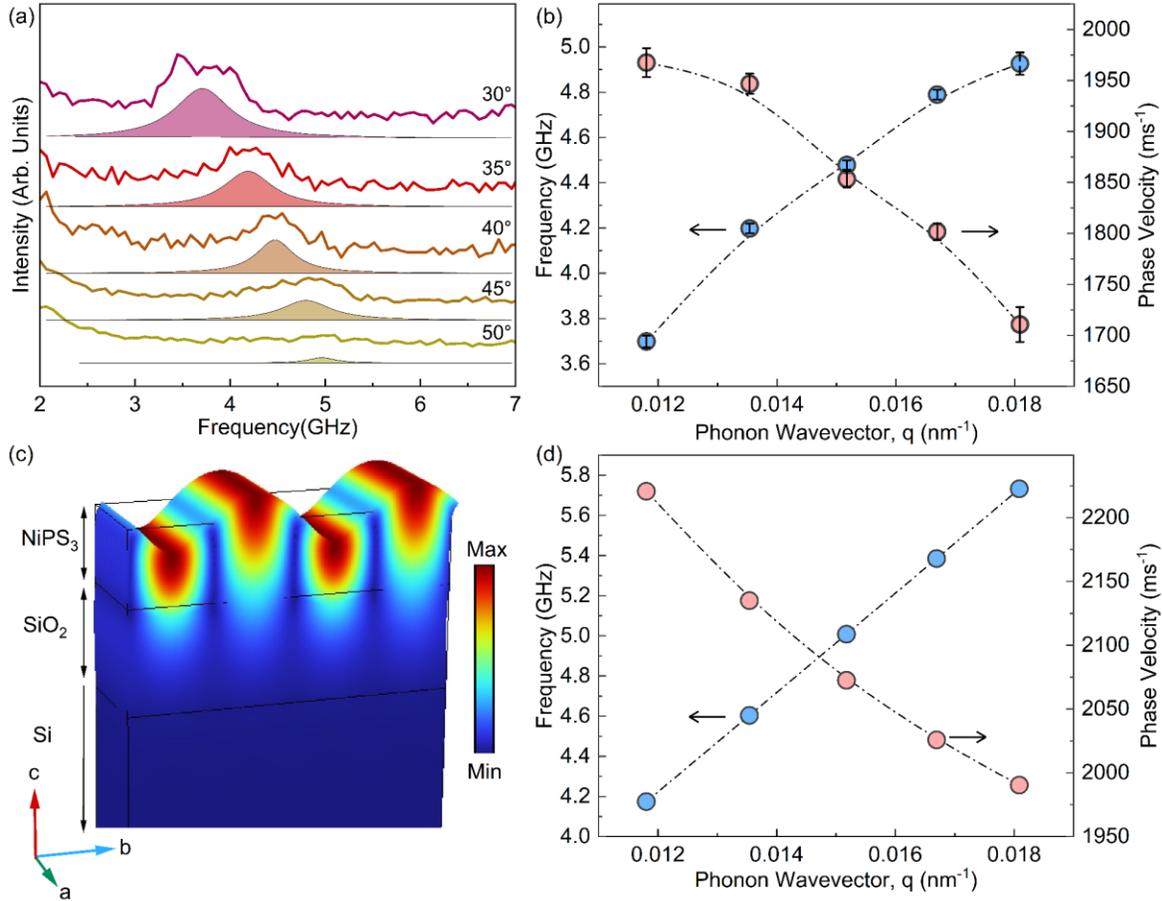

**[Fig. 3: Experimental and theoretical Rayleigh-wave dispersion in NiPS$_3$.** a) BMS spectra of the exfoliated NiPS$_3$ film collected at room temperature for incidence angles between 30° and 50°. A single dispersive peak is observed and assigned to the surface Rayleigh wave (RW). The shaded areas represent the individual Lorentzian fits used to extract the peak frequencies. (b) Spectral position, $f$, of the RW peak (blue circles) obtained by Lorentzian fitting, together with the corresponding phase velocity (red circles) as a function of wavevector, $q$. The dashed line marks the trend of decreasing phase velocity with increasing $q$. (c) Displacement field of the RW mode calculated using the continuum elastic model at $q = 0.0135$ nm$^{-1}$, showing surface-localized, elliptical motion predominantly within the NiPS$_3$ layer. (d) Simulated RW dispersion and phase velocity in the non-magnetic (NM) state, extracted from finite-element eigenmode analysis. The simulated trend shows good agreement with the experimental results in panel (b).]

The Rayleigh surface acoustic wave is governed by the elastic constants of the crystal, making its frequency highly sensitive to changes in lattice stiffness associated with magnetic ordering[44,45]. Therefore, the temperature dependence of the Rayleigh mode provides a direct probe of magnetoelastic coupling across the AFM–PM transition. To track the Rayleigh wave evolution with temperature, BMS measurements were conducted from 300 K to 80 K while holding the incidence angle fixed at $\theta = 35°$, maintaining a constant in-plane wavevector. The spectra are presented in Fig. 4 (a). Each spectrum was fit using a single Lorentzian function. As seen, the spectral position of the Rayleigh peak increases gradually upon cooling from RT to ~200 K, followed by a distinct drop at around $T_N$, and then remains nearly constant at lower temperatures. The initial increase reflects the expected reduction of anharmonic lattice vibrations upon cooling, which leads to lattice stiffening and thus a higher acoustic phase velocity. The drop in frequency at 150 K is associated with the onset of AFM order, where spin rearrangement modifies the elastic constants through magnetoelastic coupling, resulting in a measurable softening. This behavior is quantitatively reflected in the phase velocity of the Rayleigh waves, plotted in Fig. 4(b), which decreases from ~1920 ms$^{-1}$ in the PM state to ~1815 ms$^{-1}$ in the AFM state, corresponding to a ~5.5 % reduction across the transition. Once the magnetic order is established, the elastic constants stabilize, resulting in a nearly temperature-independent frequency at low temperatures.

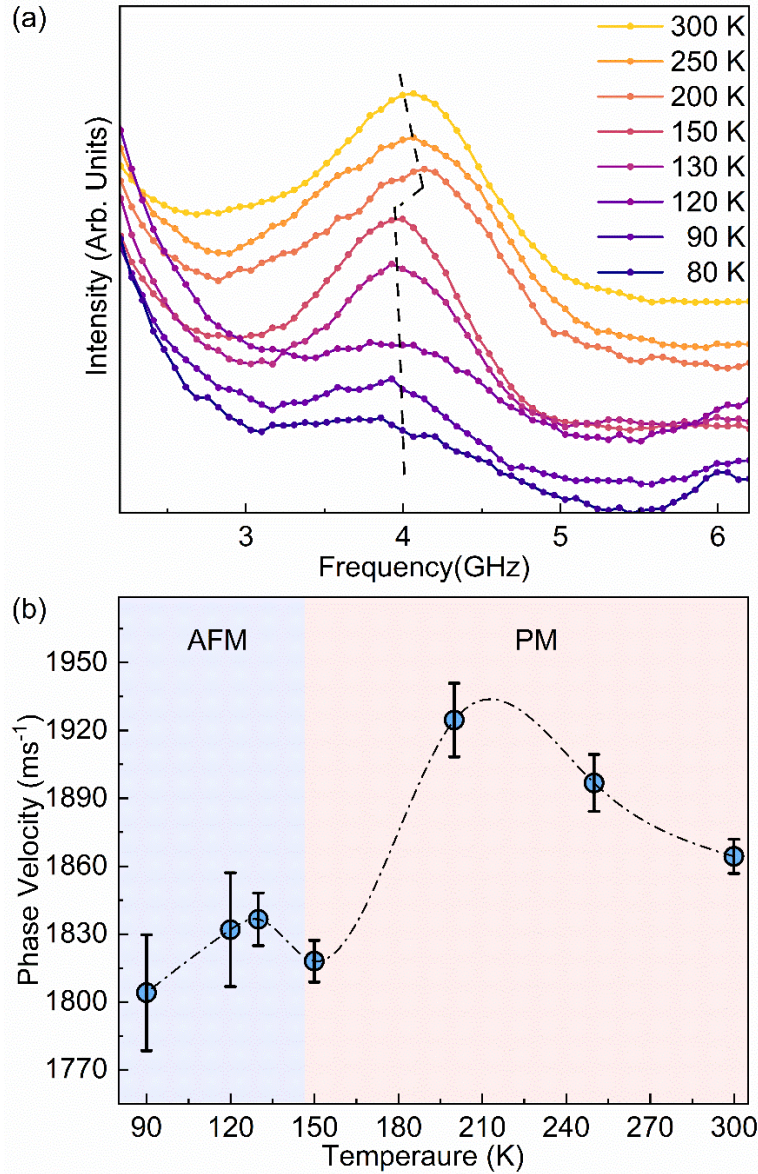

[**Fig. 4: Temperature dependence of the Rayleigh surface acoustic wave in NiPS$_3$.** (a) BMS spectra of NiPS$_3$ film exfoliated on SiO$_2$/Si substrate measured between 300 K and 80 K at constant wavevector $q = 0.0135$ nm$^{-1}$. The dashed line serves as a guide to the eye, tracking the evolution of the Rayleigh-wave peak position with temperature. Upon cooling, the Rayleigh-wave frequency initially increases due to lattice stiffening, then exhibits a pronounced drop at 150 K near the Néel temperature (~155 K), followed by saturation at lower temperatures. (b) Phase velocity of the Rayleigh-wave as a function of temperature. The velocity decreases from ~1920 ms$^{-1}$ in the paramagnetic (PM) phase to ~1815 ms$^{-1}$ in the antiferromagnetic AFM phase, corresponding to a ~5.5% reduction across the transition. The shaded regions mark the PM and AFM phases. The anomaly in velocity reflects the magnetoelastic softening associated with the onset of zigzag AFM order in NiPS$_3$.]

To further understand the origin of the Rayleigh-wave anomaly across the magnetic transition, we extended our calculation of the elastic stiffness tensor $c_{ij}$ to the AFM ground state and compared it with the previously obtained NM phase used in our simulations. The $c_{ij}$ tensors in both phases are summarized in Table 1 in standard Voigt notation. NiPS$_3$ with a monoclinic C2/m structure possesses 13 independent elastic constants. As seen in the table, several $c_{ij}$ elements undergo pronounced changes upon magnetic ordering. The most notable variation occurs in the in-plane shear stiffness $c_{66}$, which increases from 47.4 GPa in the NM phase to 55.4 GPa in the AFM phase. In contrast, interlayer and in-plane coupling terms such as $c_{12}$ and $c_{23}$ decrease, reflecting a magnetically driven softening of off-diagonal couplings. Note the sign reversal of the shear–normal coupling constant $c_{35}$, which changes from –0.5 GPa (NM) to +0.4 GPa (AFM). Because the Rayleigh surface wave involves a mixed displacement of in-plane longitudinal (LA) and out-of-plane transverse (TA) components, its frequency is governed by a complex combination of elastic coefficients rather than a single stiffness term[45]. In an anisotropic heterostructure such as NiPS$_3$/SiO$_2$/Si, additional boundary constraints and elastic mismatch further influence the effective velocity[45]. To quantify this effect, we implemented the AFM-phase elastic constants in finite-element simulations of the continuum elastic model in COMSOL Multiphysics. The calculated Rayleigh mode at $q = 0.0135$ nm$^{-1}$ shows a frequency of 4.578 GHz, corresponding to a 0.54 % decrease relative to the NM-phase value of 4.603 GHz. Although this reduction is smaller than the experimentally observed drop (~6 %), the trend confirms that the AFM ordering leads to elastic softening of the Rayleigh mode. The smaller calculated change likely arises because DFT captures both AFM and NM phases at 0 K, whereas finite-temperature effects further renormalize the elastic response in experiments. Additionally, the PM phase was approximated assuming NM configuration, which neglects spin fluctuations and thus introduces additional uncertainty into the comparison.

[**Table 1.** Elastic constant $c_{ij}$ (GPa) of NiPS$_3$ obtained from DFT calculation.]

| | $c_{11}$ | $c_{22}$ | $c_{33}$ | $c_{12}$ | $c_{13}$ | $c_{23}$ | $c_{44}$ | $c_{55}$ | $c_{66}$ | $c_{15}$ | $c_{25}$ | $c_{35}$ | $c_{46}$ |
|---|---|---|---|---|---|---|---|---|---|---|---|---|---|
| NM | 152.3 | 148.4 | 44.8 | 50.7 | 9.0 | 12.3 | 9.7 | 9.9 | 47.4 | 3.4 | -4.2 | -0.5 | -2.5 |
| AFM | 150.2 | 151.2 | 41.3 | 39.5 | 8.4 | 9.4 | 9.7 | 10.1 | 55.4 | 2.1 | -1.9 | 0.4 | -2.1 |

From an atomistic perspective, the contrast between the weak, mode-selective frequency shifts of optical phonons and the pronounced softening of the Rayleigh surface acoustic wave reflects two distinct forms of spin–lattice coupling. Optical phonons involve *out-of-phase* atomic displacements within the unit cell that locally perturb super-exchange pathways[20]. Their magnetic renormalization is governed by the derivative of the exchange integral with respect to an atomic displacement $u_\alpha$, such that the optical phonons frequency shift obeys $\Delta\omega_{opt} \propto (\partial J_{ij}/\partial u_\alpha) < \mathbf{S_i} \cdot \mathbf{S_j} >$ [15,19,46]. Here, $J_{ij}$ is the exchange interaction between spins $S_i$ and $S_j$. Because different bonds contribute with opposite signs of $\partial J_{ij}/\partial u_\alpha$, these effects are strongly mode-dependent and can yield either softening or hardening across the transition [15,25]. In contrast, acoustic and Rayleigh waves correspond to *in-phase*, long-wavelength displacements that generate a slowly varying strain field $\varepsilon_{\alpha\beta} = (1/2)[(\partial u_\alpha/\partial x_\beta) + (\partial u_\beta/\partial x_\alpha)]$ [Ref. [47]]. This coherently modulates all exchange interactions through $J_{ij} = J_{ij}^{(0)} + (\partial J_{ij}/\partial \varepsilon_{\alpha\beta})\varepsilon_{\alpha\beta}$, giving rise to a magnetoelastic energy density $E_{me} = -\sum_{\alpha\beta} B_{\alpha\beta} \varepsilon_{\alpha\beta} S_\alpha S_\beta$, where $B_{\alpha\beta}$ quantifies the collective derivative $\partial J_{ij}/\partial \varepsilon_{\alpha\beta}$ [Ref. [48]]. When long-range AFM order sets in, this coupling renormalizes the elastic tensor as $c_{\alpha\beta}^{eff} = c_{\alpha\beta}^{(0)} - \lambda_{\alpha\beta}^2 \chi_m$, with $\chi_m$ the magnetic susceptibility and $\lambda_{\alpha\beta} \propto B_{\alpha\beta}$. Consequently, the Rayleigh-wave velocity, $v_R \propto \sqrt{c_{eff}/\rho}$ decreases by $\Delta v_R/v_R \simeq 5.5\%$, implying an effective modulus reduction of $\Delta c_{eff}/c_{eff} \simeq (2)(\Delta v_R/v_R) \simeq 11\%$. This quantitatively links the macroscopic elastic softening to the same exchange–striction mechanisms responsible for optical phonon renormalization, but now integrated coherently over the entire lattice. The observed Rayleigh anomaly thus represents the *continuum limit* of spin–phonon coupling, where the collective modulation of exchange by uniform strain yields a measurable renormalization of the elastic constants across the magnetic transition.

The substantial softening of the Rayleigh-wave velocity across the AFM–PM transition demonstrates that the acoustic elastic response of NiPS$_3$ is strongly renormalized by magnetic ordering. Because NiPS$_3$ is highly anisotropic and the Rayleigh mode in the NiPS$_3$/SiO$_2$/Si heterostructure depends on a complex combination of elastic constants, we performed a systematic sensitivity analysis by individually increasing each of the 13 independent $c_{ij}$ values by 20 % in the continuum elastic model while keeping all others fixed (see Supplementary Table 2). The calculated response revealed that the Rayleigh-wave velocity is most sensitive to $c_{33}$ and $c_{44}$, with weaker contributions from $c_{22}$, $c_{23}$, and $c_{46}$, whereas the remaining elastic constants have a negligible impact. The experimentally observed ~6 % drop in phase velocity at $T_N$ therefore implies a sizable reduction in the shear- and out-of-plane-related elastic coefficients, due to a strong magnetoelastic renormalization triggered by the emergence of zigzag antiferromagnetic order. These results show that the spin-driven reconfiguration of the elastic tensor in NiPS$_3$ is sufficiently large to measurably alter the propagation of surface acoustic phonons, revealing a giant spin–lattice interaction rarely reported in two-dimensional magnets. Our findings are in agreement with a recent theoretical work reporting that NiPS$_3$ exhibits an exceptionally large spin–phonon coupling constant, an order of magnitude higher than most other 2D magnets such as FePS$_3$, CrBr$_3$, or Cr$_2$Ge$_2$Te$_6$, placing NiPS$_3$ among the strongest spin–phonon–coupled materials[49–52]. The Rayleigh-wave anomaly observed here provides a direct long-wavelength spectroscopic manifestation of this strong spin–phonon coupling, establishing NiPS$_3$ as a benchmark platform for studying magnetoelastic interactions and phonon-mediated functionalities in low-dimensional antiferromagnets.

## Conclusions

In summary, we have demonstrated that long-wavelength acoustic phonons provide a uniquely sensitive probe of spin–lattice coupling in the quasi-two-dimensional antiferromagnet NiPS$_3$. Our Brillouin–Mandelstam measurements reveal a pronounced softening of the Rayleigh surface mode across the Néel temperature, corresponding to ~5.5% reduction in phase velocity. The latter directly demonstrates that the onset of zigzag AFM order renormalizes the long-wavelength elastic constants. First-principles calculations of the elastic tensor, together with continuum modelling, corroborate the magnetic-order–induced changes in elasticity and link them to exchange-striction at the microscopic level. These results establish the continuum limit of spin–phonon interactions in NiPS$_3$, where microscopic exchange interactions coherently reshape the macroscopic stiffness of the lattice. By revealing how magnetic order governs the elastic energy landscape, this work provides a foundation for lattice-controlled spin phenomena and offers opportunities for magnetically tunable phononic, magnonic, and strain-mediated spintronic devices in low-dimensional materials.

## Methods

**Raman Spectroscopy Measurements:** Raman spectroscopy measurements were performed on exfoliated NiPS$_3$ film using a Renishaw inVia Qontor system with a λ = 532 nm excitation laser over the temperature range from 77 K to 300 K. The measurements were conducted in a backscattering geometry. The scattered light was collected under different polarization configurations to distinguish the vibrational symmetries of the phonon modes. The spectrometer was calibrated before each measurement using an undoped SiO$_2$/Si wafer, with the Si peak fixed at 520 cm$^{-1}$. A low laser power of 0.024 mW was used to ensure a high signal-to-noise ratio while minimizing laser-induced heating of the sample.

**Brillouin-Mandelstam Spectroscopy Measurements:** The Brillouin-Mandelstam spectroscopy measurements were performed in a backscattering geometry using a continuous-wave solid-state diode-pumped laser (Spectra-Physics, Excelsior, USA) at the excitation wavelength of λ = 532 nm. The scattered light was directed into a high-resolution, high-contrast 3 + 3 tandem Fabry-Perot interferometer (The Table Stable, Switzerland), coupled to a detector and spectrum analyzer.

**Density Functional Theory (DFT) Calculations:** All first-principles calculations were performed within the framework of density functional theory (DFT) using the Vienna *ab initio* Simulation Package (VASP)[53–55]. The plane-wave energy cutoff was set to 600 eV. Gaussian smearing method with a smearing width of 0.01 eV was used. The exchange–correlation functional was treated using the generalized-gradient approximation in the PBE form[56], augmented with Grimme's D3 correction to account for long-range van der Waals interactions[57]. Structural relaxations were performed with force and total-energy convergence thresholds of 0.005 eV/Å and 1×10$^{−9}$ eV, respectively. Elastic constants were obtained from stress–strain relations using finite differences with a strain amplitude of 0.015, and four displacements. A Γ-centered *k*-point mesh of 3×2×3 was employed for Brillouin zone sampling. The simulations were carried out for a 1×4×2 supercell of NiPS$_3$ (lattice dimensions (AFM): 11.499 × 19.914 × 12.308 Å), consistent with its monoclinic (C2/m) symmetry. All calculations were performed for both the AFM ground state and the NM (paramagnetic proxy) state, ensuring full relaxation in the respective magnetic configuration prior to elastic tensor evaluation. Phonon dispersion and density of states in the AFM and NM states are shown in Supplementary Fig. 8 and Supplementary Fig. 9.)

**Finite Element Simulations:** Finite-element simulation was performed by using COMSOL Multiphysics with the Solid Mechanics Module. The system is modeled with the small-strain, linear electrodynamic equation in the frequency domain, $-\rho\omega^2 u = \nabla \cdot s$, where $u$ is the displacement vector, $\rho$ is the density, $\omega$ is angular frequency, and $s$ is the stress tensor related to the strain by the stiffness tensor $c_{ij}$. Simulations were carried out in 3D by applying an eigenfrequency study to reflect the experimental excitation of the system. The geometric parameters used in the simulations were aligned with the actual characteristics of the samples studied experimentally. The model consisted of a rectangular block of NiPS$_3$ layer of thickness 220 nm on top of a SiO$_2$ layer (300 nm) over a Si half-space, represented numerically by a finite thickness. Floquet periodic boundary conditions (PBC) were applied along the x and y axes of the unit cell with a phase shift $u(x + L_x) = u(x)e^{ikL_x}$, where $u(x)$ is the displacement along x direction, $L_x$ is the model length, and $k$ is the periodic phase set to the experimentally studied phonon wavevector, q. These PBCs were implemented on both faces of the unit cell to ensure consistent values of the elastic tensor components and density throughout the entire modelled structure. The top surface of NiPS$_3$ is set to be free and the bottom surface of the Si was fixed. NiPS$_3$ was treated as a linear elastic, anisotropic solid.

## Acknowledgements

This work was supported, in part, by the National Science Foundation (NSF), Division of Materials Research (DMR) *via* Project No. 2436557 entitled "Controlling Electron, Magnon, and Phonon States in Quasi-2D Antiferromagnetic Semiconductors for Enabling Novel Device Functionalities" awarded to F.K. and A.A.B. The instrumentation used in this project was developed with the support of NSF provided to A.A.B. and F.K. *via* a Major Research Instrument (MRI) DMR Project No. 2019056 entitled "Development of a Cryogenic Integrated Micro-Raman-Brillouin-Mandelstam Spectrometer." The density-functional-theory computations were supported, in part, by resources provided by the Auburn University Easley Cluster.

## Author Contributions

F.K. and A.A.B. coordinated the project and provided funding support. F.K. led the experimental and theoretical data analysis and manuscript preparation. A.A.B. contributed to data analysis. Z.E.N. conducted the BMS measurements and COMSOL simulations and contributed to data analysis. M.S.H.B. performed temperature- and polarization-dependent Raman spectroscopy and contributed to data analysis. V.K. carried out the DFT calculations under the supervision of K.K. M.Z. conducted Raman measurements. D.W. assisted with BMS and Raman measurements and contributed to data analysis. All authors contributed to manuscript preparation.